\documentclass[a4paper]{spie}
\usepackage[]{graphicx}
\usepackage{amsmath}
\usepackage{amsfonts}
\usepackage{amssymb}

\newcommand{\be}{\begin{equation}}
\newcommand{\ee}{\end{equation}}
\newcommand{\bea}{\begin{eqnarray}}
\newcommand{\eea}{\end{eqnarray}}
\newcommand{\ba}{\begin{array}}
\newcommand{\ea}{\end{array}}

\title{Activation process driven by strongly non-Gaussian noises} 

\author{Bart{\l}omiej Dybiec and Ewa Gudowska--Nowak
\skiplinehalf
Marian~Smoluchowski Institute of Physics, Jagellonian University, Reymonta~4, 30--059~Krak\'ow, Poland\\
}

\authorinfo{Further author information:\\B.D.: E-mail: bartek@th.if.uj.edu.pl, Telephone: +48 (12) 663 5566, Fax: +48 (12) 633 40 79 \\ E.G--N.: E-mail: gudowska@th.if.uj.edu.pl, Telephone: +48 (12) 663 5567, Fax: +48 (12) 633 40 79 
}

\begin{document} 
\maketitle 

\begin{abstract}
The constructive role of non-Gaussian random fluctuations is studied in the context of the passage over the dichotomously switching potential barrier. Our attention focuses on the interplay of the effects of independent sources of fluctuations: an additive stable noise representing non-equilibrium external random force acting on the system and a fluctuating barrier. In particular, the influence of the structure of stable noises on the mean escape time and on the phenomenon of resonant activation (RA) is investigated. By use of the numerical Monte Carlo method it is documented that the suitable choice of the barrier switching rate and random external fields may produce resonant phenomenon leading to the enhancement of the kinetics and the shortest, most efficient reaction time. 
\end{abstract}

\keywords{Stochastic resonance, resonant activation, escape time, numerical evaluation of the resonant activation, stable random variables}

\section{INTRODUCTION}\label{sect:intro} 
The problem of noise induced phenomena is of special interest in non-equilibrium natural systems covering broad class of examples ranging from physics and chemistry to biological sciences.\cite{lefever} In particular, systems driven by both, external noises and thermal fluctuations have been shown to manifest several counterintuitive phenomena leading to e.g. the enhancement of the detection of weak sensory signal or optimization of a directed transport.\cite{gammaitoni} Among various types of ``resonant behaviors'', so called resonant activation (RA) has attracted a special attention as a feature problem describing kinetics in fluctuating environments subject to nonequilibrium fluctuations.\cite{doe,boguna,iwaniszewski,dybiec1,dybiec2} The phenomenon is detected as a most efficient, shortest passage over a fluctuating barrier with a maximum resonant activation being a function of the correlation time of the barrier noise. The typical theoretical models analyzing the RA phenomena are usually based on a Langevin equation approach assuming the overdamped limit.\cite{gammaitoni,doe,dybiec1,dybiec2} Accordingly, the influence of the external thermal bath of the surroundings on a Brownian particle is described in such an equation by time-dependent random force which is commonly assumed to be represented by a white Gaussian noise. That postulate is compatible with the assumption of a short correlation time of fluctuations, much shorter than the time-scale of the macroscopic motion and assumes that weak interactions with the bath lead to independent random variations of the parameter describing the motion. In more formal, mathematical terms Gaussianity of the state-variable fluctuations is a consequence of the Central Limit Theorem which states that normalized sum of independent and identically distributed random variables with finite variance converges to the Gaussian probability distribution. If, however, after random collisions jump lengths are ruled by broad distributions leading to the divergence of the second moment, the statistics of the process changes significantly. The existence of the limiting distribution is then guaranteed by the generalized L\'evy-Gnedenko\cite{janicki} limit theorem. According to the latter, normalized sums of independent, identically distributed random variables with infinite variance converge in distribution to the L\'evy statistics. At the level of the Langevin equation, L\'evy noises are generalization of the Brownian motion and describe results of strong collisions between the test particle and the surrounding environment. In this sense, they lead to different models of the bath that go beyond a standard ``close-to-equilibrium'' Gaussian description.\cite{dietlevsen,bartek,bartek2} 

In the recent studies\cite{dybiec2} we have shown that the same kind of behavior which was observed in the systems coupled to the sources of Gaussian fluctuations can be observed when instead of equilibrated or close to equilibrium heat reservoirs a non-equilibrated thermal baths are considered. On a single particle level, the contact with non-equilibrated thermal bath can be described by a generalized Langevin equation, i.e. Langevin equation in which instead of Gaussian fluctuations some heavy tailed L\'evy-type fluctuations are incorporated.\cite{fogedby} Here we report some basic results for resonant activation processes driven by L\'evy flights\cite{bartek,bartek2} and compare the appearance of the phenomenon for a Gaussian and for a heavy-tailed $\alpha$-stable noise. Furthermore, we design a strategy that allows to determine the feature of underlying noises and recognition of their characteristic properties. In Section~\ref{sect:model} the archetypal model and basic mathematical introduction are presented. More formal definitions and necessary technical information are moved to the Appendix. Section~\ref{sect:results} contains results of numerical analysis. The discussion and final remarks are presented in Section~\ref{sect:discussion}.

\section{MODEL} \label{sect:model}
\begin{figure}
\begin{center}
\begin{tabular}{c}
\includegraphics[height=7cm]{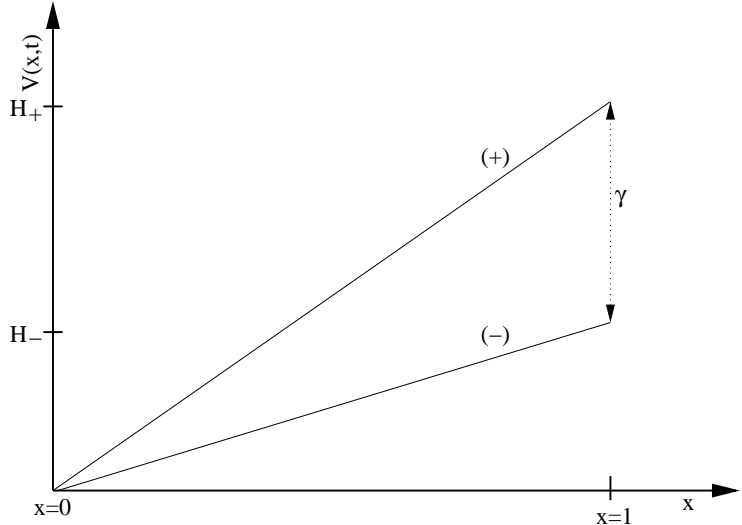}
\end{tabular}
\end{center}
\caption{ \label{fig:model} 
A model of the potential barrier studied in the paper. The barrier height fluctuates dichotomously between two barrier configurations characterized by two height $H_{\pm}$, in this presentation $H_-=0,H_+=8$. A particle starts its diffusive motion, in $t=0$, at a reflecting boundary, $x=0$, and continues until the absorption at the absorbing boundary, $x=1$, at $t=\tau$.}
\end{figure}

We study the motion of an overdamped Brownian particle moving in a potential field between reflecting $x=0$ and absorbing $x=1$ boundaries in the presence of noise that modulates the barrier height. Time evolution of a state variable $x(t)$ is given in terms of the generalized Langevin equation 
\be
\frac{dx}{dt} = -V'(x)+g\eta(t)+\sqrt{2}\zeta(t) = -V_\pm^{'}(x)+\sqrt{2}\zeta(t),
\label{lang}
\ee
where prime means differentiation over $x$, $\zeta(t)$ is a white L\'evy process\cite{janicki} originating from the contact with non-equilibrated bath and $\eta(t)$ represents a Markovian dichotomous noise of intensity $g$ taking one of two possible values $\pm 1$. Both $\zeta$ and $\eta$ noises are assumed to be statistically independent and autocorrelation of the dichotomous noise is equal to 
$\langle (\eta(t)-\langle\eta\rangle)(\eta(t')-\langle\eta\rangle)\rangle=\exp(-2\gamma |t-t'|)$. 
For simplicity, a particle mass, a friction coefficient and the Boltzman constant are all set to 1. The time-dependent potential $V_\pm(x)$ is represented by a linear barrier with its height switching between two values $H_\pm$ with an average rate $\gamma$ (Fig.~\ref{fig:model})
\be
V_\pm(x)=H_\pm x, \qquad g=\frac{H_{-}-H_{+}}{2}.
\ee

Equivalent to the stochastic ordinary differential equation~(\ref{lang}) is a deterministic fractional Fokker-Planck equation\cite{fogedby} (FFPE) for the distribution function 
\bea
\frac{\partial p_\pm(x,t)}{\partial t} = \frac{\partial}{\partial x} \left[ \frac{\partial V_\pm(x)}{\partial x}p_\pm(x,t) \right] + D\nabla^\alpha p_\pm(x,t) 
 + \gamma p_\mp(x,t)-\gamma p_\pm(x,t).
\label{sffpe}
\eea
In the above FFPE $p_\pm(x,t)$ are probability density functions for finding a particle at time $t$ in the vicinity of $x$, while potential takes the value $V_\pm(x)$. The fractional derivative $\nabla^\alpha$ is understood in the sense of the Fourier Transform\cite{jespersen,metzler} 
\be
\nabla^\alpha=-\int\frac{dk}{2\pi}e^{ikx}|k|^\alpha,
\ee
with $\alpha=2$ corresponding to the standard Brownian diffusion. The coefficient $D$ represents the generalized diffusion coefficient with the dimension\cite{fogedby} $[D]=cm^{\alpha}sec^{-1}$. The initial and boundary conditions for Eq.~(\ref{sffpe}) are
\be
p_\pm(x,t)|_{t=0}=\frac{1}{2}\delta(x),\;\;\;\;p_\pm(x,t)|_{x=1}=0.
\label{sffpecondition}
\ee
The first condition represents equal chances of finding initially a potential barrier in any of two possible configurations $(\pm)$. The second condition is implied by the absorbing boundary, which is located at $x=1$. Furthermore, as mentioned earlier, at $x=0$ reflecting boundary is located.

In the approach presented herein, instead of solving Eq.~(\ref{sffpe}), information on the system are drawn from the statistics of numerically\cite{janicki,mannella} generated trajectories satisfying the generalized Langevin equation~(\ref{lang}). At a single trajectory level sampled from the Monte Carlo\cite{newman} study of the problem, the initial condition for Eq.~(\ref{lang}) is
\be
x(t)|_{t=0}=0,
\label{conddiscrete}
\ee
i.e. initially particle is located, with probability equal to 1, at the reflecting boundary and potential barrier is, with probability $1/2$, in any of two possible configurations $(\pm)$. Eq.~(\ref{conddiscrete}) is a discrete analogue of the condition (\ref{sffpecondition}) for the FFPE~(\ref{sffpe}).

The quantity under consideration is the mean first passage time ($\mathrm{MFPT}$), i.e. the average time that particle spends in the system before it becomes absorbed. The information 
about the $\mathrm{MFPT}$ is drawn from the statistics of numerically generated trajectories satisfying the generalized Langevin equation (\ref{lang}), i.e. the examined $\mathrm{MFPT}$ is estimated as a first moment of the distribution $f(\tau)$ of first passage times (FPT),\footnote{In the presentation $t$ and $\tau$ are used. The $\tau$ is used when $\mathrm{MFPT}=\langle\tau\rangle$ is discussed or some remarks on FPTs are expressed. The $t$ is used in the context of integration variable or in any context associated with the survival probability $G(t)$, which is constructed by analysis of FPTs, $\tau$. This distinction is provided to distinguish time, $t$, which can take any positive values, from the FPT, $\tau$, which is a discrete random variable taking also positive values only.} obtained from the ensemble of simulated realizations of the stochastic process in question. More precisely, FPT is calculated as time $\tau$ for which a Brownian particle starting initially from $x=0$ reaches $x=1$ for the first time and $\mathrm{MFPT}$ is estimated as $\langle \tau \rangle$. 

Based solely on the statistics of generated trajectories we estimate the mean
first passage time as

\be
\mathrm{MFPT}=\langle\tau\rangle=\int_0^1dx\int_0^\infty[p_-(x,t)+p_+(x,t)]dt=\int_0^\infty G(t)dt.
\ee
where $G(t)=\int_0^1[p_-(x,t)+p_+(x,t)]dx$ is a survival probability, i.e. the probability that at the time $t$ a Brownian particle is still in the system, i.e. in the interval $[0,1)$. Survival probability function, $G(t)$, is easy to construct from sample realizations of the process given by Eq.~(\ref{lang}). The function $G(t)$ could provide method for identification of underlying stochastic process $\zeta(t)$ in the cases when methods based on examination of $\mathrm{MFPT}$ fail to provide insight into underlying dynamics. Furthermore, from $G(t)$ it is possible to calculate $\mathrm{MFPT}$ and to construct $f(\tau)$.

\section{RESULTS} \label{sect:results}
In the standard Gaussian scenario phenomena expressing constructive role of the noise are examined as a function of the noise intensity $\sigma^2$, which is related to temperature. For the additive white Gaussian noise, $\sigma^2$ which is equal to the variance of the Gaussian distribution, is the only one noise parameter. In the Gaussian regime stochastic resonance takes place when for a given value of $\sigma^2$ examined quantity (e.g. signal to noise ratio, spectral amplification) is maximal.\cite{gammaitoni} For the non-Gaussian noises situation is more subtle. It is caused by the fact that stable distributions, which are considered here, are characterized by four parameter family ($\alpha,\beta,\sigma,\mu$). The stability index $\alpha$ ($\alpha\in(0,2]$), skewness parameter $\beta$ ($\beta\in[-1,1]$), scaling parameter $\sigma$ ($\sigma\in\mathbb{R^+}$) and location parameter $\mu$ ($\mu\in\mathbb{R}$). The stability index describe how heavy tailed distribution is, i.e. PDFs asymptotically behave like $|\zeta|^{-(1+\alpha)}$. The $\beta$ parameter describe skewness of the distribution, i.e. which part of the distribution is ``heavier'', e.g. for $\alpha<1$ and $\beta=1$ the distribution is totally skewed and random variable $L_{\alpha,1}(x;\sigma,\mu)$ takes only values greater than $\mu$. The scale parameter $\sigma$ scales the distribution width. All of those parameters determine how likely large fluctuations can be. Finally, the $\mu$ parameter is responsible for the location of the modal value.

In the following simulations value of $\mu$ has been set to 0
and $\sigma=1/\sqrt{2}$. Such a choice of $\sigma$ reconstruct standard normal distribution $N(0,1)$ for $\alpha=2$, i.e. the well known Langevin equation with the additive Gaussian white noise is recovered and comparison with previous studies of RA\cite{doe,dybiec1,dybiec2} can be performed. Remaining parameter, $\alpha$, has taken values from the whole allowed range, however very small $\alpha$, $\alpha<0.2$, has not been investigated. Furthermore, due to numerical instability $\alpha=1$ was excluded from the analysis.
The MC simulation has been performed for $\alpha$ increasing by $0.1$ from simulation to the simulation. Simulation for other values of parameters were also performed and results have been presented elsewhere.\cite{bartek2}

In the presented study, for a given barrier setup, various types of $\alpha$-stable noises were investigated. The potential barrier was switching between different heights $H_-=0$ and $H_+=8$. Such a choice corresponds to the changing of the potential barrier between the barrier and no barrier, i.e. diffusion in a potential field and free Brownian motion. Specific values of $H_\pm$ correspond to the previous study of RA\cite{dybiec1,dybiec2,bartek} and allow for simple comparison with previously calculated $\mathrm{MFPTs}$ for the considered model in the presence of Gaussian, Cauchy and L\'evy-Smirnoff noises.\cite{bartek}

\begin{figure}
\begin{center}
\begin{tabular}{c}
\includegraphics[height=7cm]{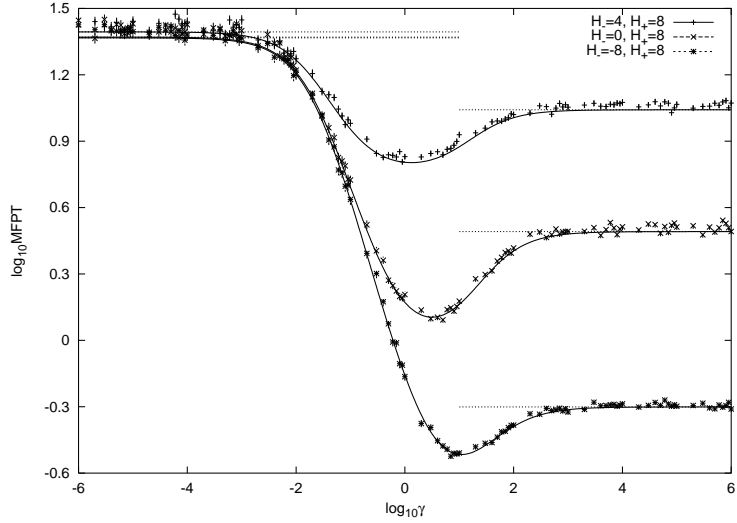}
\end{tabular}
\end{center}
\caption{ \label{fig:gauss} 
MFPT$(\gamma)$ for linear potential barriers switching between different heights $H_\pm$: `$+$': $H_-=4,\;H_+=8$; `$\times$': $H_-=0,\;H_+=8$; `$\ast$': $H_-=-8,\;H_+=8$. The driving noise is the Gaussian noise, which is the special case of the L\'evy stable noise, i.e. L\'evy stable noise with the stability index $\alpha=2$ and the scale parameter $\sigma=1/\sqrt{2}$. Solid lines represent exact solution constructed by direct integration of the backward Fokker--Planck equation.\cite{doe,dybiec1} Numerical results were obtained by use of Monte Carlo simulation of Eq.~(\ref{lang}) with time step $dt=10^{-4}$ and averaged over $N=10^3$ realizations. Error bars represent deviation from the mean and usually remain within the symbol size. Additionally, exact asymptotic values of $\mathrm{MFPT}$ are presented, i.e values of $\mathrm{MFPT}(\gamma\to 0)$ and $\mathrm{MFPT}(\gamma\to\infty)$.}
\end{figure} 

\begin{figure}
\begin{center}
\begin{tabular}{c}
\includegraphics[height=7cm]{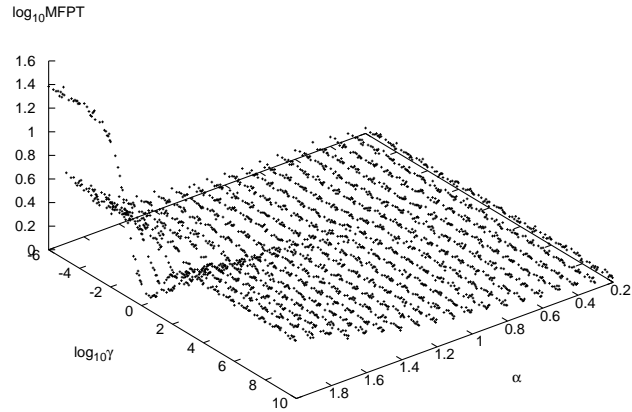}
\end{tabular}
\end{center}
\caption{ \label{fig:alpha08} 
MFPT$(\alpha,\gamma)$ for $\beta=0,\sigma=1/\sqrt{2}$ for the linear potential barrier switching between $H_-=0,\;H_+=8$. The results were calculated by direct integration of Eq.~(\ref{lang}) with the time step $dt=10^{-4}$ and averaged over $N=10^3$ realizations.}
\end{figure}

\begin{figure}
\begin{center}
\begin{tabular}{c}
\includegraphics[height=7cm]{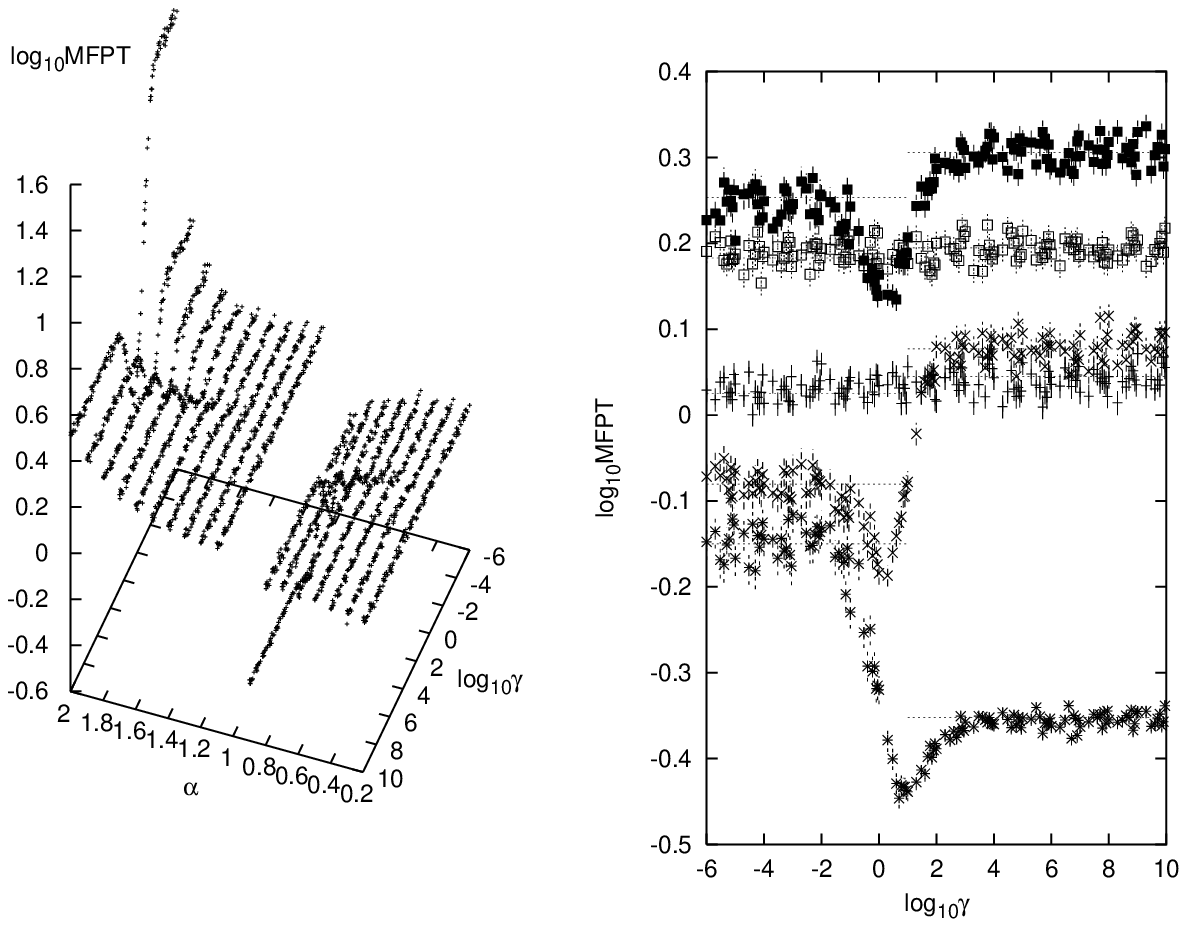}
\end{tabular}
\end{center}
\caption{ \label{fig:alphaskewed08b} 
$\mathrm{MFPT}(\alpha,\gamma)$ for $H_-=0,\;H_+=8$ and $\beta=1,\sigma=1/\sqrt{2}$ (left panel) and sample cross-sections $\mathrm{MFPT}(\gamma)$ for various $\alpha$: `$+$': $\alpha=0.2$; `$\times$': $\alpha=0.8$; `$\ast$': $\alpha=0.9$; `$\square$': $\alpha=1.1$ and `$\blacksquare$': $\alpha=1.7$ (right panel). The results were calculated by direct integration of Eq.~(\ref{lang}) with the time step $dt=10^{-4}$ and averaged over $N=10^3$ realizations. Additionally, asymptotic values of $\mathrm{MFPT}$ are presented, i.e values of $\mathrm{MFPT}(\gamma\to 0)$ and $\mathrm{MFPT}(\gamma\to\infty)$. Asymptotic values of $\mathrm{MFPT}$ were calculated using Monte Carlo methods with $dt=10^{-4}$ and averaged over $N=5\times10^3$ realizations. Due to numerical problems results for $\alpha=1$ are not presented.}
\end{figure}

\begin{figure}
\begin{center}
\begin{tabular}{c}
\includegraphics[height=7cm]{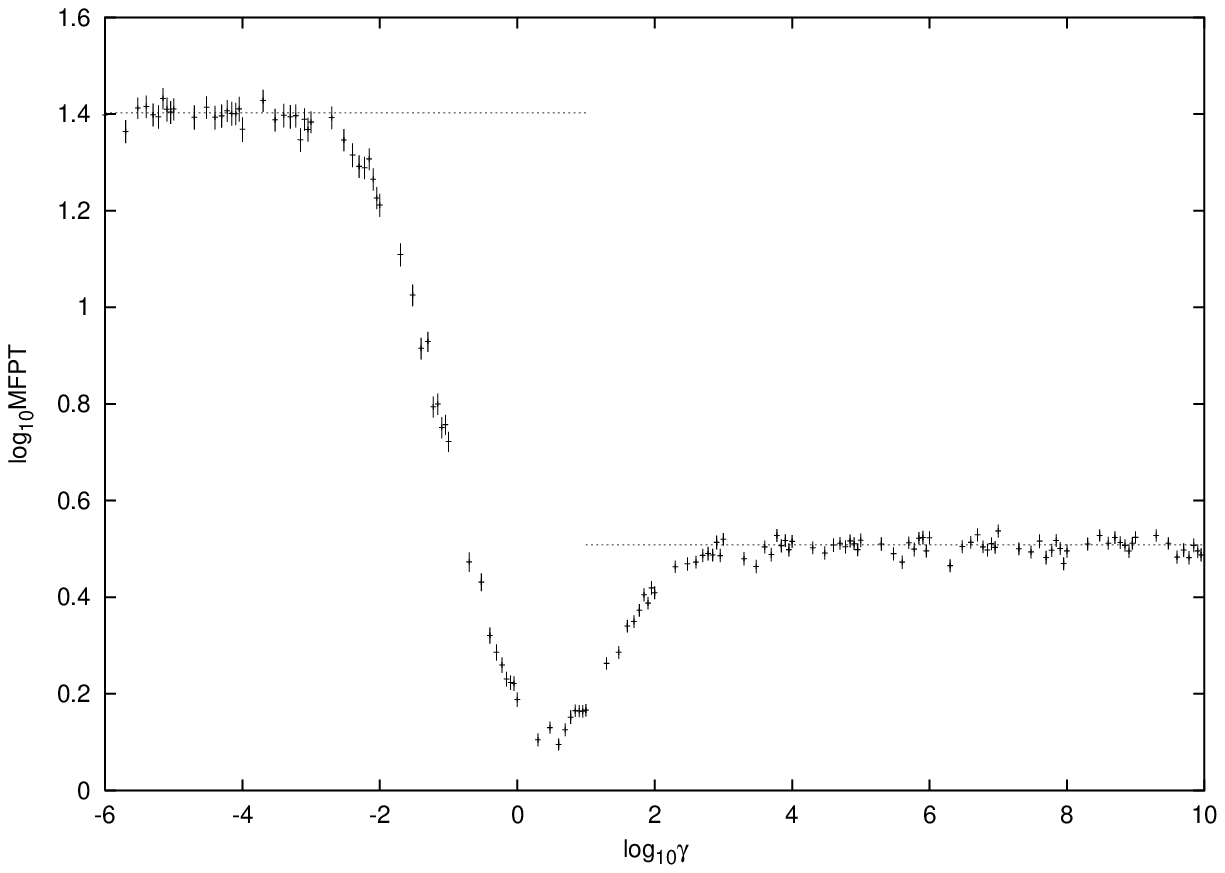}
\end{tabular}
\end{center}
\caption{ \label{fig:alpha08gaus} 
Sample cross section of the $\mathrm{MFPT}(\alpha,\gamma)$ surface (left panel of Fig.~\ref{fig:alphaskewed08b}; $H_-=0,\;H_+=8$ and $\beta=1,\sigma=1/\sqrt{2}$) with asymptotic values of $\mathrm{MFPT}$, for $\alpha=2$, i.e. for the case when additive fluctuations are of the Gaussian type.
}
\end{figure}

\begin{figure}
\begin{center}
\begin{tabular}{c}
\includegraphics[height=7cm]{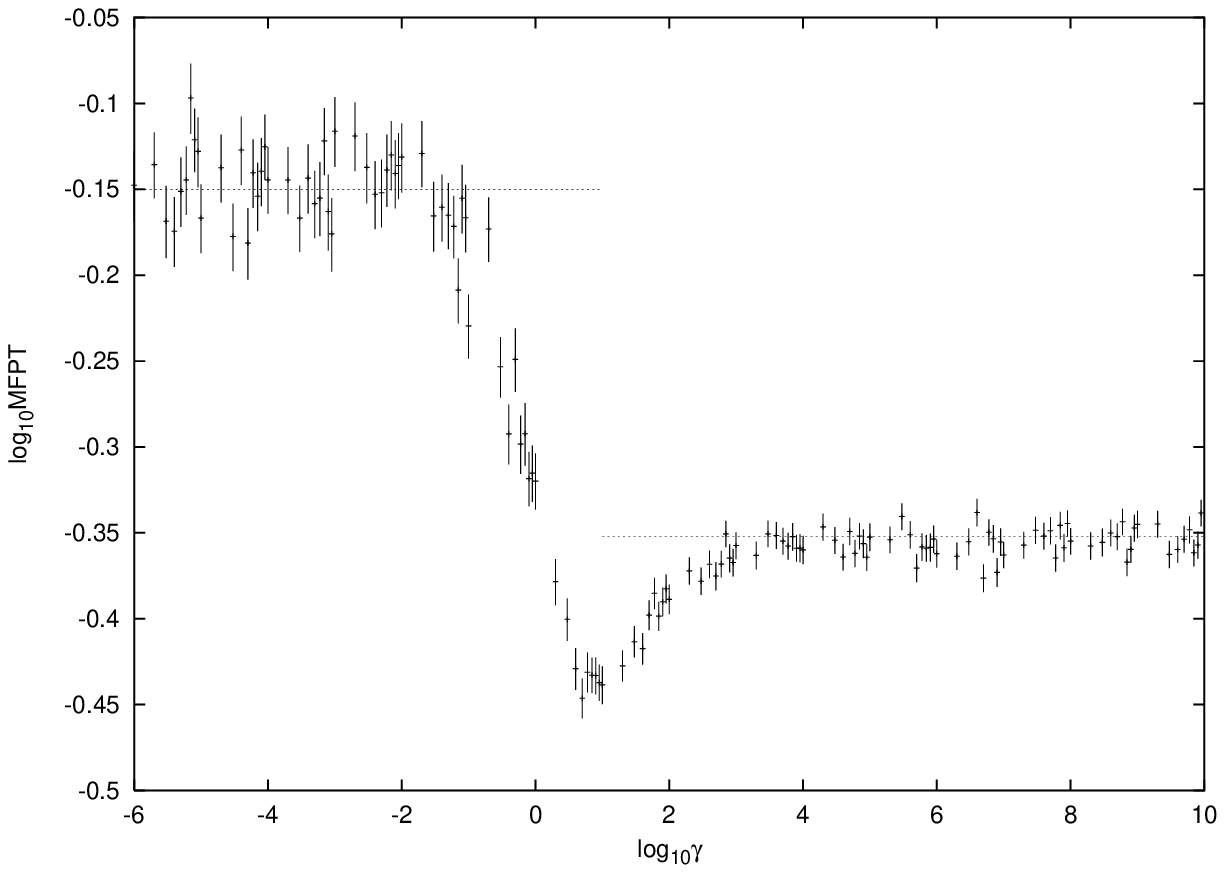}
\end{tabular}
\end{center}
\caption{ \label{fig:alphaskewed0809mfpt} 
Sample cross section of the $\mathrm{MFPT}(\alpha,\gamma)$ surface (left panel of Fig.~\ref{fig:alphaskewed08b}; $H_-=0,\;H_+=8$ and $\beta=1,\sigma=1/\sqrt{2}$) with asymptotic values of $\mathrm{MFPT}$, for $\alpha=0.9$, i.e. for the case when additive fluctuations are of a heavy-tailed type.}
\end{figure}

\begin{figure}
\begin{center}
\begin{tabular}{cc}
\includegraphics[width=7cm,height=7cm]{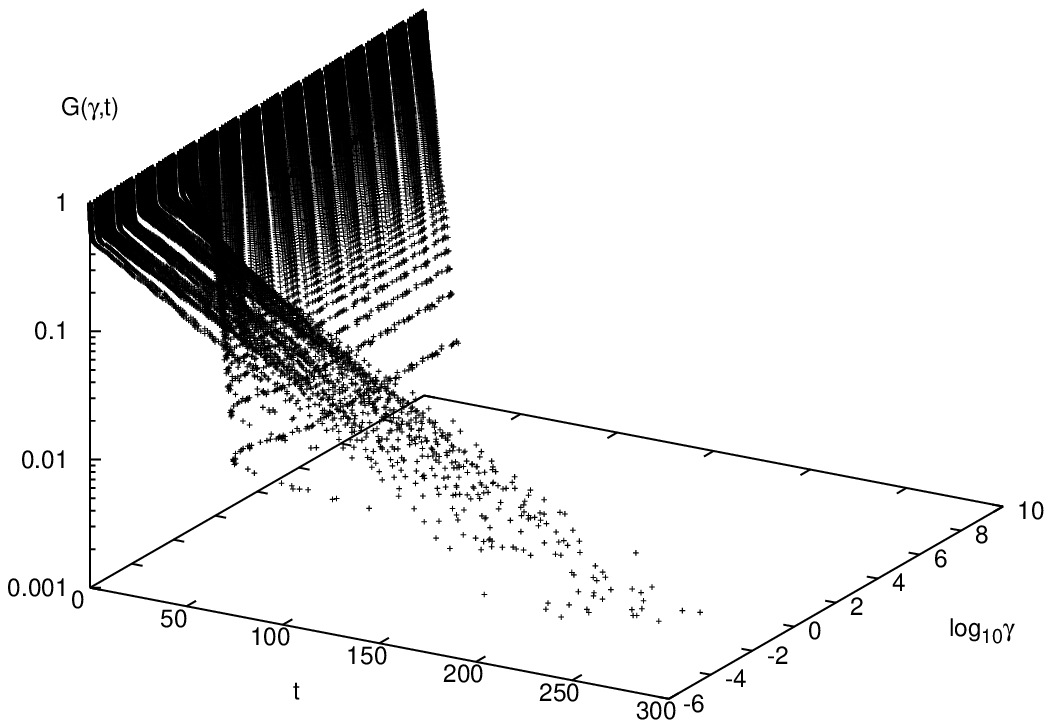} & \includegraphics[width=7cm,height=7cm]{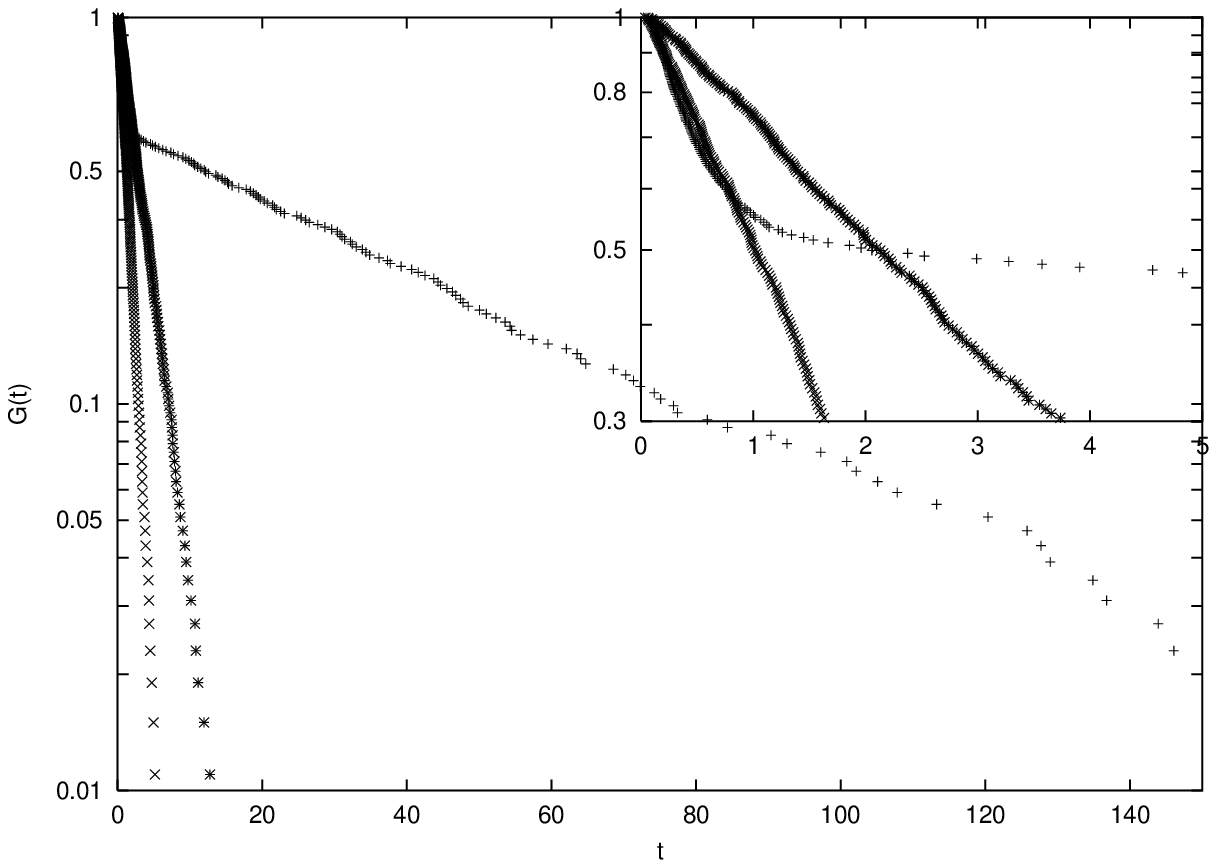}
\end{tabular}
\end{center}
\caption{ \label{fig:alpha08gausssurv} 
Survival probability distribution, $G(\gamma,t)$, (left panel) corresponding to the $\mathrm{MFPT}(\gamma)$ curve for $\alpha=2,\beta=1,\sigma=1/\sqrt{2}$ with $H_-=0,H_+=8$ (cf. Fig.~\ref{fig:alpha08gaus}) and sample cross-sections (right panel) of $G(t)$ for various $\gamma$: `$+$': $\log_{10}\gamma=-6$ (low frequency); `$\times$': $\log_{10}\gamma=0.48$ (resonant frequency) and `$\ast$': $\log_{10}\gamma=10$ (high frequency). The results were calculated by direct integration of Eq.~(\ref{lang}) with the time step $dt=10^{-4}$ and averaged over $N=10^3$ realizations. Note for log scale: z-axis (left panel) and y-axis (right panel). The inset in the right panel represent behavior of the survival probability, $G(t)$, for small $t$.
}
\end{figure}

\begin{figure}
\begin{center}
\begin{tabular}{cc}
\includegraphics[width=7cm,height=7cm]{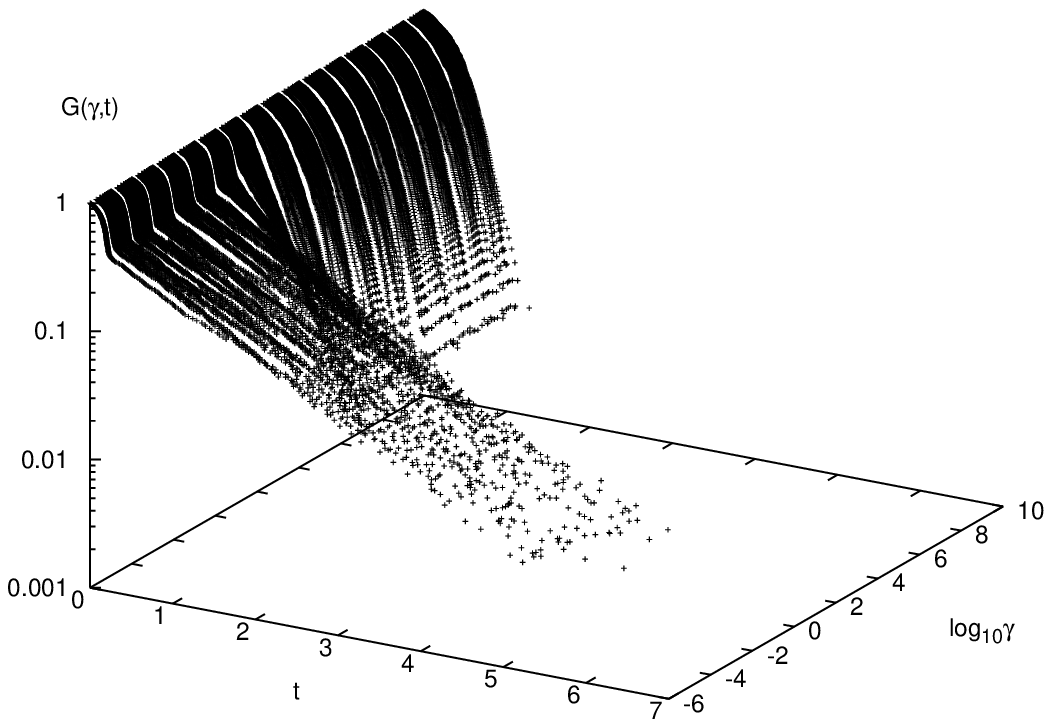} & \includegraphics[width=7cm,height=7cm]{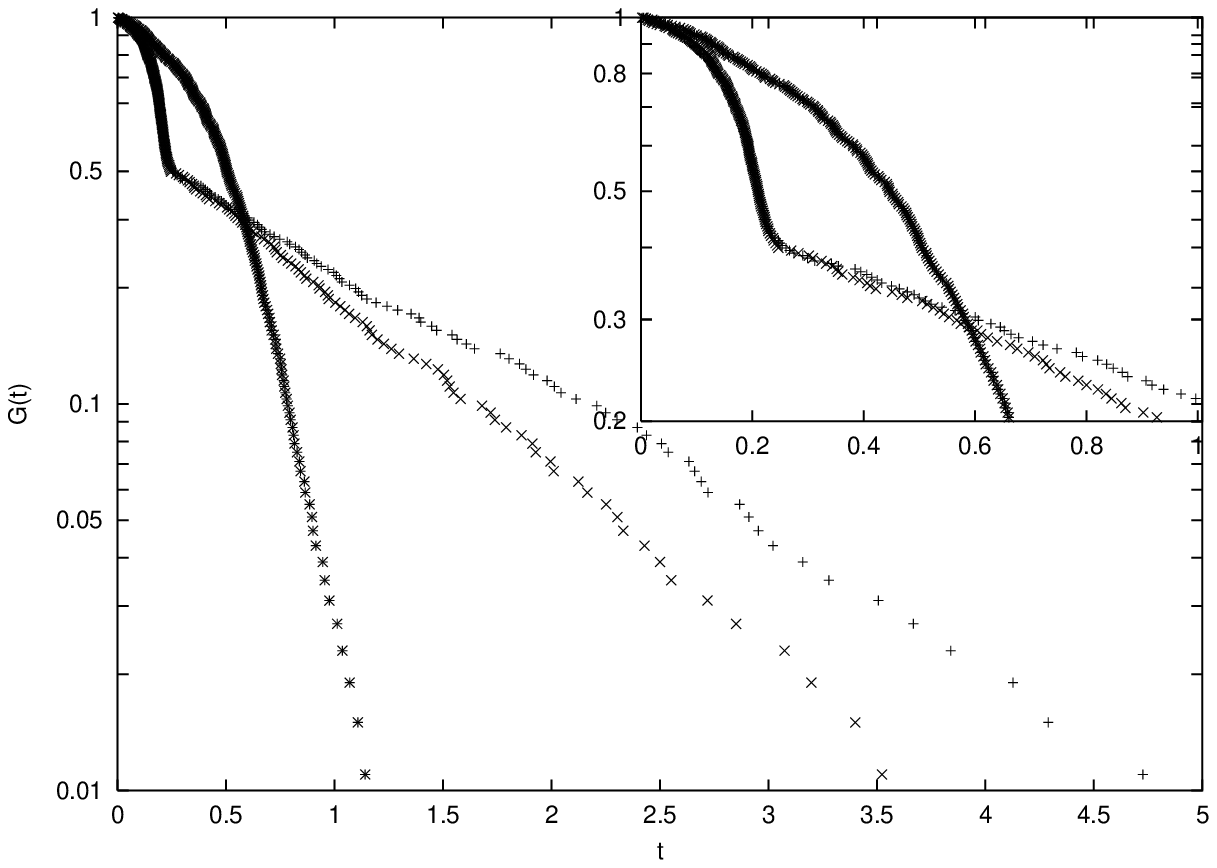}
\end{tabular}
\end{center}
\caption{ \label{fig:alphaskewed0809surv} 
Survival probability distribution, $G(\gamma,t)$, (left panel) corresponding to the $\mathrm{MFPT}(\gamma)$ curve for $\alpha=0.9,\beta=1,\sigma=1/\sqrt{2}$ with $H_-=0,H_+=8$ (cf. Fig.~\ref{fig:alphaskewed08b}) and sample cross-sections of $G(t)$ (right panel) for various $\gamma$: `$+$': $\log_{10}\gamma=-6$ (low frequency); `$\times$': $\log_{10}\gamma\approx1$ (resonant frequency) and `$\ast$': $\log_{10}\gamma=10$ (high frequency). The results were calculated by direct integration of Eq.~(\ref{lang}) with the time step $dt=10^{-4}$ and averaged over $N=10^3$ realizations. Note for log scale: z-axis (left panel) and y-axis (right panel). The inset in the right panel represent behavior of the survival probability, $G(t)$, for small $t$.
}
\end{figure}

For the testing purposes of the implemented numerical procedures MC solutions of Eq.~(\ref{lang}) for $\alpha=2$ were compared with exact solutions of backward Fokker Planck Equation.\cite{doe,dybiec1} The obtained numerical results are in the perfect agreement with the exact solutions.\cite{doe,dybiec1} Error bars which are shown in Fig.~\ref{fig:gauss} represent deviation from the mean and were calculated by standard MC error analysis method.\cite{newman} Numerical results presented in the Fig.~\ref{fig:gauss} were calculated by direct integration\cite{mannella} of Eq.~(\ref{lang}), with respect to the general $\alpha$-stable measure, with the time step $dt=10^{-4}$ and averaged over $N=10^3$ sample realizations. The same time step, $dt$, and ensemble of $N$ trajectories were used to analyze other cases under the study, i.e. other types of stable noises (Figs.~\ref{fig:alpha08}-\ref{fig:alphaskewed0809surv}). Different time steps, number of repetitions were examined, but they do not changed results quantitatively. For example, results calculated for $dt=10^{-5}$ are the same like for the larger time step, $dt=10^{-3}$ or $dt=10^{-4}$, except the $\alpha=1,\beta\ne0$ cases, which for that reason were excluded. Presented results are divided into two groups. The first group correspond to the symmetric, $\beta=0$, $\alpha$-stable noises (Figs.~\ref{fig:gauss} and~\ref{fig:alpha08}). The second group to the totally skewed, $\beta=1$, stable noises. 

By examination of Figs.~\ref{fig:alpha08} and~\ref{fig:alphaskewed08b} (left panel) it can be seen that the typical Gaussian behavior of $\mathrm{MFPT}$ curves changes with the change of the stability index $\alpha$. For symmetric $\alpha$-stable noises, $\beta=0$ (Fig.~\ref{fig:alpha08}) with decreasing of the stability index, $\alpha$, minima become less distinct and finally for small $\alpha$ phenomena of the RA vanishes. For totally skewed $\alpha$-stable noises situation is more complicated and less regular. Obviously for $\alpha=2$ the Gaussian limit is recovered and the RA is well visible. With decreasing stability index $\alpha$ minima become less pronounced and for $\alpha\approx 1.5$ the RA vanishes. For $\alpha=0.9$ the RA is visible again and disappears for $\alpha<0.7$ (right panel of Fig.~\ref{fig:alphaskewed08b}). 

From Figs.~\ref{fig:alphaskewed08b}--\ref{fig:alphaskewed0809mfpt} it can be concluded that the same kind of behavior, i.e. a typical non-monotonic shape of $\mathrm{MFPT}$ curves, is observed in two opposite limits, i.e. for equilibrium Gaussian fluctuations and for non-equilibrated fluctuations described by heavy tailed, asymmetric stable noise. Observation of MFPT curves only (Figs.~\ref{fig:alpha08gaus} and \ref{fig:alphaskewed0809mfpt}) do not provide any information about underlying noise. Therefore more detailed study is required. In Figs.~\ref{fig:alpha08gausssurv} and \ref{fig:alphaskewed0809surv} survival probabilities, $G(\gamma,t)$, are presented in a semi-log scale. Right panels present sample cross sections of $G(\gamma,t)$ surface, for various values of $\gamma$: low, high and resonant switching frequency.

In the case when underlying fluctuations are Gaussian (Fig.~\ref{fig:alpha08gausssurv}) it can be concluded that $G(\gamma,t)$ for high values of $\gamma$ is distributed exponentially while for small values of $\gamma$ two distinct slopes are visible. These slopes correspond to the various time regimes responsible for passages over barrier in $(\pm)$ configurations.\cite{dybiec2} For large values of the $\gamma$ survival probability distribution is exponential and particle escapes through average potential barrier.\cite{dybiec2} For non-Gaussian fluctuations (Fig.~\ref{fig:alphaskewed0809surv}) $G(\gamma,t)$ is no longer distributed exponentially. Here again two time scales of the escape process over a fluctuating potential barrier are visible for small $\gamma$ and one for large values of $\gamma$. Inspection of insets demonstrates different character of $G(t)$ for Gaussian (Fig.~\ref{fig:alpha08gausssurv}) and non-Gaussian statistics (Fig.~\ref{fig:alphaskewed0809surv}) with noticeable difference in shape caused by different underlying fluctuations. Therefore methods based on the survival probability analysis can be applied for recognition of the shape of the underlying fluctuations.

\section{DISCUSSION} \label{sect:discussion}
In this study, we have investigated motion of a Brownian particle in a potential field subjected to Markovian dichotomous fluctuations in the presence of external, heavy tailed fluctuations arising from the contact with a non-equilibrated thermal bath. The potential field has been approximated by a linear potential barrier. For such a model phenomena of resonant activation have been investigated. Furthermore we focused on method of recognition of underlying noise type.

For the numerical study we applied numerical procedures that allow to generate family of strictly stable L\'evy variables\cite{janicki} and in consequence allow to perform integration of stochastic differential equations with respect to the family of strictly stable measures. Using that procedures resonant activation and survival probability for various types of stable noises were examined. In particular it was checked that for $\alpha=2$, the well studied Gaussian case is recovered. Monte Carlo results for Gaussian noises constructed by general procedure, for simulation of the generalized Langevin equation are in perfect agreement with previous numerical\cite{dybiec1,dybiec2} and analytical results.\cite{doe} For easier comparison with results of previous studies value of $\sigma$ was set to $1/\sqrt{2}$.

For nonsymmetric stable noises, for $H_-=0,H_+=8$, for very small values of the stability index $\alpha$ RA is not visible. Increasing value of the $\alpha$ RA appears. Furthermore with increase of $\alpha$ phenomena of RA vanishes and again for $\alpha$ approaching to 2 is visible, $\alpha=2$ is the well known Gaussian case. The fact that the shape of $\mathrm{MFPT}$ curve for heavy tailed, skewed distributions (e.g. $\alpha=0.9,\beta=1$) arising from the contact with not equilibrated bath is similar to the $\mathrm{MFPT}$ curve in the presence of the Gaussian noise ($\alpha=2$) arising from the contact with equilibrated thermal bath is very interesting and makes the problem of recognition of the underlying noise more complicated. It is very intriguing that similar behavior is recovered in two opposite limits. The analysis of $\mathrm{MFPT}$ only does not allow to recognize the character of underlying noise. The examination of survival probability, $G(t)$, or FPTs distribution, $f(\tau)$, provide a deeper insight into the problem and can be useful in making any statement about underlying thermodynamics. However, analysis of $G(t)$ provide more information about the system, it is not always sufficient. Therefore further studies are required.

Like in the Gaussian regime\cite{doe}, in the every case under the study the MFPT for small $\gamma$ is equal to the average value of MFPTs for both barrier configurations, i.e. $\mathrm{MFPT}(\gamma\to 0)=\frac{1}{2}[\mathrm{MFTP}(H_-)+\mathrm{MFTP}(H_+)]$. For large $\gamma$ the MFPT is equal to the MFPT over the average potential barrier, i.e. $\mathrm{MFPT}(\gamma\to \infty)=\mathrm{MFTP}\left(\frac{1}{2}(H_-+H_+)\right)$.

\appendix \label{sect:math}
\section{Stable Noises and their Representations}
Let $\{X_1,X_2,X\}$ are some independent random variables distributed according to the same unknown probability distribution. The most intuitive definition\cite{janicki} of the stable variable, can be written as
\be
aX_1+bX_2\stackrel{\mathrm{d}}{=}cX+d,
\label{def}
\ee
where $\stackrel{\mathrm{d}}{=}$ denotes equality in a distribution sense, i.e. the $\alpha$-stable variables are random variables for which the sum of random variables is distributed according to the same distribution as each variable. Real constants $c,\;d$ in Eq.~(\ref{def}) allow for rescaling and shifting of the initial probability distribution. From Eq.~(\ref{def}) the characteristic function of $\alpha$-stable random variables can be determined.\cite{metzler} The characteristic function of the stable distribution can be parameterized in various ways. In the strictly stable parameterization\cite{janicki,weron} $L_{\alpha,\beta}(\zeta;\sigma,\mu)$, a characteristic function of the L\'evy type variables is given by 
\be
\phi(k)= 
\left\{
\begin{array}{ll}
\exp\left[ -\sigma^\alpha|k|^\alpha\left( 1-i\beta\mbox{sign}(k)\tan
\frac{\pi\alpha}{2} \right) +i\mu k \right], & \mbox{for}\;\;\alpha\neq 1, \\
\exp\left[ -\sigma|k|\left( 1+i\beta\frac{2}{\pi}\mbox{sign} (k) \ln|k| \right) + i\mu k \right], & \mbox{for}\;\;\alpha=1, \\
\end{array}
\right.
\label{charakt}
\ee
with $\alpha\in(0,2],\; \beta\in[-1,1],\; \sigma\in(0,\infty),\; \mu\in(-\infty,\infty)$ 
and $\phi(k)$ defined in the Fourier space
\be
\phi(k) = \int d\zeta e^{-ik\zeta} L_{\alpha,\beta}(\zeta;\sigma,\mu).
\ee
Parameter $\alpha$ is called the stability index, $\beta$ describes skewness of the distribution, $\sigma$ is responsible for its scaling and $\mu$ is a location parameter. Generally, for $\beta=\mu=0$ PDFs are symmetric while for $\beta=\pm1$ and $\alpha\in(0,1)$ they are totally skewed, i.e. $\zeta$ is always positive only or negative only depending on the sign of skewness parameter $\beta$.\cite{janicki} 
Despite the fact that stable probability distributions asymptotically behave like $\propto |\zeta|^{-(\alpha+1)}$ analytical expression for their probability distributions are known only in the few cases. For $\alpha=2$ and any $\beta$ resulting distribution is Gaussian
\be
L_{2,0}(\zeta;\sigma,\mu)=\frac{1}{2\sigma\sqrt{\pi}} \exp\left(-\frac{(\zeta-\mu)^2}{4\sigma^2} \right)\equiv\mathit{N}(\mu,2\sigma^2),
\ee
however by the matter of tradition $\beta=0$ is usually chosen. For $\alpha=1,\beta=0$ Cauchy distribution is obtained
\be
L_{1,0}(\zeta;\sigma,\mu)=\frac{\sigma}{\pi}\frac{1}{(\zeta-\mu)^2+\sigma^2},
\ee
and for $\alpha=0.5,\beta=1$ L\'evy-Smirnoff distribution
\be
L_{1/2,1}(\zeta;\sigma,\mu) = \left( \frac{\sigma}{2\pi}
\right)^{\frac{1}{2}}(\zeta-\mu)^{-\frac{3}{2}} \exp\left(-\frac{\sigma}{2(\zeta-\mu)} \right).
\ee
For $L_{\alpha,\beta}(\zeta;\sigma,\mu)$distributions moments of order $\alpha$ exist, i.e. the integral $\int_{-\infty}^{\infty}L_{\alpha,\beta}(\zeta;\sigma,\mu)\zeta^\alpha d\zeta$ is finite. This result in the conclusion that for every $\alpha$ variance of stable distribution does not exist and for $\alpha\le1$ the average value also does not exist.

Position of the Brownian particle is calculated by direct integration of Eq.~(\ref{lang}) with respect to the $\alpha$-stable measure $L_{\alpha,\beta}(s)$
\bea
x(t) & =& -\int_{t_0}^{t}\left[ V'(x(s))-g\eta(s)\right] ds + \int_{t_0}^{t} dL_{\alpha,\beta}(s) \nonumber \\
&= & -\int_{t_0}^{t}V_\pm^{'}(x(s))ds+\int_{t_0}^{t}dL_{\alpha,\beta}(s).
\label{lcalka}
\eea
The $L_{\alpha,\beta}$ measure in Eq.~(\ref{lcalka}) can be approximated by\cite{janicki,dietlevsen}
\bea
\int_{t_0}^{t}f(s)dL_{\alpha,\beta}(s) & \approx & 
\sum\limits_{i=0}^{N-1}f(i\Delta s)M_{\alpha,\beta}\left([i\Delta s,(i+1)\Delta s)\right) \nonumber \\
& \stackrel{\mathrm{d}}{=} & \sum\limits_{i=0}^{N-1}f(i\Delta s)\Delta
s^{1/\alpha}\varsigma_i,
\label{stochasticintegral}
\eea
where $\varsigma_i$ is distributed according to the $L_{\alpha,\beta}(\varsigma;\sigma,\mu=0)$, $N\Delta s=t-t_0$ and $M_{\alpha,\beta}([i\Delta s,(i+1)\Delta s))$ is the measure of the interval $[i\Delta s,(i+1)\Delta s)$.

Stable random variables $\varsigma$, corresponding to the characteristic function (\ref{charakt}), can be generated according to the following recipes.\cite{janicki,weron} For $\alpha\neq1$ it is necessary to calculate 
\be
\varsigma = D_{\alpha,\beta,\sigma} \frac{\sin(\alpha(V+C_{\alpha,\beta})) }{ (\cos V)^{\frac{1}{\alpha}}} 
\left[
\frac{\cos(V-\alpha(V+C_{\alpha,\beta}))}{W}
\right]^{\frac{1-\alpha}{\alpha}},
\label{recipe1}
\ee
with constants $B,C,D$ given by
\be
C_{\alpha,\beta}=\frac{\arctan\left(\beta\tan(\frac{\pi\alpha}{2})\right)}{
\alpha},
\ee
\be
D_{\alpha,\beta,\sigma}=\sigma\left[ \cos\left(
\arctan\left(\beta\tan(\frac{\pi\alpha}{2})\right) \right) \right]^{-\frac{1}{\alpha}}.
\ee
For $\alpha=1$, $\varsigma$ can be obtained from the formula
\be
\varsigma = \frac{2\sigma}{\pi} \left[ (\frac{\pi}{2}+\beta V)\tan(V) -\beta\ln \left( \frac{\frac{\pi}{2}W\cos V}{\frac{\pi}{2}+\beta V} \right) \right] + \mu.
\label{recipe2}
\ee
In the above equations $V$ and $W$ are independent random variables. $V$ is uniformly distributed in the interval $(-\frac{\pi}{2},\frac{\pi}{2})$ and $W$ is exponentially distributed with a unit mean.\cite{janicki,weron}

In the simulation, Eq.~(\ref{lcalka}) is numerically examined till $x<1$. When $x\ge 1$ for the first time the recorded value of $t=\tau$ is an FPT for the problem under the study. From the ensemble of FPTs, $\tau$, the $\mathrm{MFPT}$, $\langle \tau \rangle$, is estimated and the survival probability, $G(t)$, is constructed. Survival probability, $G(t)$, is connected with the cumulative distribution of the FPTs
\be
F(\tau)=1-G(t)|_{t=\tau}.
\ee
Furthermore, from $G(t)$ the probability density of FPTs, $f(\tau)$,\cite{dybiec2,gardiner} can be calculated
\be
f(\tau)=-\frac{d}{dt}G(t)|_{t=\tau}.
\ee

\acknowledgments 
The Authors acknowledge the financial support from the Polish State Committee for Scientific Research (KBN) through grants 1P03B06626 and 2P03B08225.


\end{document}